\pdfoutput=1

\documentclass[aps,prl,10pt,twocolumn,showpacs,superscriptaddress]{revtex4-1}

\usepackage{latexsym}
\usepackage{graphics,epstopdf}
\usepackage{newlfont}
\usepackage{amssymb}
\usepackage{amsfonts}
\usepackage{amsmath}
\usepackage{amsthm}
\usepackage{graphicx}
\usepackage{epsfig}
\usepackage{color}
\usepackage{bm}

\usepackage{times}

\usepackage{amsmath,epsfig,graphicx,amssymb}
\usepackage[latin1]{inputenc}
\usepackage[margin=1in]{geometry}
\usepackage{amsmath}
\usepackage{amsfonts}
\usepackage{amssymb}
\usepackage{makeidx}
\usepackage{graphicx}
\usepackage{setspace}
\newcommand{\beq}{\begin{equation}}
\newcommand{\eeq}{\end{equation}}
\newcommand{\ben}{\begin{eqnarray}}
\newcommand{\een}{\end{eqnarray}}
\newcommand{\bes}{\begin{subequations}}
\newcommand{\ees}{\end{subequations}}
\newcommand{\bFig}{\begin{figure}}
\newcommand{\eFig}{\end{figure}}
\newcommand{\ket}[1]{|{#1}\rangle}
\newcommand{\bra}[1]{\langle{#1}|}
\newcommand{\braket}[2]{\langle {#1} | {#2} \rangle}
\newcommand{\ketbra}[2]{| \rangle {#1}  {#2} \langle |}
\renewcommand{\t}[1]{\textrm{#1}}
\newcommand{\com}[1]{[\textsc{#1}]}

\begin{document}

%\title{The quantum state could be mere information}
\title{An Epistemic Model of Quantum State with Ontic Probability Amplitude}

\author{Arun Kumar \surname{Pati}}
\email{akpati@hri.res.in}
\affiliation{Quantum Information and Computation Group\\
Harish-Chandra Research Institute, Chhatnag Road, Jhunsi, 
Allahabad, India\\}

\author{Partha \surname{Ghose}}
\affiliation{Centre for Astroparticle Physics and Space Science (CAPSS), Bose Institute, Block EN, Sector V, Salt Lake, Kolkata 700 091, India\\}

\author{A. K. \surname{Rajagopal}}
\affiliation{Inspire Institute Inc., Alexandria, VA, USA and Quantum Information and Computation Group\\
Harish-Chandra Research Institute, Chhatnag Road, Jhunsi, 
Allahabad, India }

\begin{abstract}
We first prove that ontological models of the quantum state which are capable of reproducing the Born 
probability rule and fall in the class of $\psi$-epistemic models are inconsistent with the Sch{\"o}dinger time evolution. We then model the ontic state space as
a complex projective Hilbert space that embeds the projective Hilbert space of quantum mechanics and define a minimalist epistemic state as an average over a set of ``hidden states'' in the larger space. We show that such a model incorporates probability amplitudes and admits an epistemic interpretation of 
quantum states. 
Finally, we prove a second theorem to show that such a model is compatible with locality but ontic models are not. 
\end{abstract}

\maketitle
\vskip 0.3in

Although quantum mechanics has reigned as an outstandingly successful and accurate description of the physical world for almost a century, the interpretation of 
its state function has been of considerable debate since its inception. Central to this debate have been entanglement, measurement and violations of Bell-CHSH inequalities 
by the quantum state, signatures which have been hailed as its hallmarks \cite{bell,bell01,bell02}. Some have advocated a realist interpretation while others have preferred a 
subjective or epistemic interpretation. The most imperative question, then, is whether the wavefunction is an objective entity which is determined by the elements of 
reality or is a state of knowledge about the underlying reality. 

Recently a no-go theorem has been proved by Pusey-Barrett-Rudolph (PBR) \cite{pbr} with a couple of reasonable assumptions to rule out a subjective (epistemic) 
interpretations of the quantum state.
 In another work, it has been shown, under the assumption of free-choice of measurement settings, that only a realist or ontic interpretation of the wavefunction is
 possible \cite{cr}. Lewis {\em et al} \cite{pg} have, however, shown that if one drops the preparation independence assumption and also slightly weakens the 
definition of an epistemic state, it is possible to have an epistemic interpretation of quantum states. Using continuity and a weak separability 
assumption, Patra, Pironio and Massar \cite{patra} have argued that epistemic states are incompatible with quantum theory. However, 
the situation is far from clear and continues to attract physicists \cite{rws,rws1,sf,lei,hardy,gr}.

In this letter, we will first show (in Theorem-I below) that certain ontological models that reproduce the Born probability rule are 
%ruled out because they are 
inconsistent with the Schr\"{o}dinger evolution. Ontological models that are $\psi$-epistemic fall in this class. This is a simple and powerful demonstration
 of all previous results that aim to rule out $\psi$-epistemic models. Our proof holds without any additional assumption such as preparation independence, free-choice 
of measurement settings, or weak separability. 

Notwithstanding this, then, we ask whether it is still possible to have an 
epistemic interpretation of quantum state since it has been argued that such an interpretation may be preferable on many counts \cite{fuchs}. We show that 
if one suitably modifies the way probabilistic predictions of quantum theory are reproduced in the ontic description, it is possible to retain its epistemic nature. 
Towards that aim we introduce a certain structure of the ontic state space and a suitable definition of quantum states. To be specific, we postulate that the ontic
state space is a complex projective Hilbert space which embeds the complex projective Hilbert space of quantum mechanics, and define a quantum state as an average over 
a small dense range of unknown ontic states--which we call ``hidden-states''-- in this larger space. We emphasize that we do not use a probability distribution and an 
indicator function (response function) to reproduce Born's probability rule as is usually done. We directly relate quantum states to ontic states through a probability amplitude.
 This is a major departure from all the ontological models that exist in the literature.
In some sense, such quantum states are close to the ontic states but somewhat smeared. This may be thought of as a ``minimalist $\psi$-epistemic'' model which does not
 depart too much from reality and yet can resolve many paradoxical features of the quantum world. Furthermore, we prove a second theorem to show that such an epistemic 
model is compatible with locality but ontic models are not. 

In order to have a clear perspective of ontological models and their implications, it is important to state at the outset some useful notions defined by 
Harrigan and Spekkens \cite{hs} (henceforth referred to as HS) that are being followed in the current literature. First, an `ontological model' is set in the background 
of an `operationally' defined theory whose primitives are preparation and measurement procedures. The goal of an operationally defined theory is to prescribe the 
probabilities of different outcomes of measurements, given various preparation procedures. An `ontological model' of an operational theory is one whose primitives are properties 
of microscopic systems. In such a model a preparation procedure is assumed to prepare a system with certain properties, and measurements are supposed to reveal something
 about these properties. An `ontic state' in such a model is defined as a complete prescription of the properties of a system, and is denoted by $\lambda$. The space of 
such ontic states is denoted by $\Lambda$. It is assumed that even when an observer knows the preparation procedure $P$, she may not know the exact ontic state that is 
produced by this preparation procedure, and assigns over $\Lambda$  
a probability distribution $\mu(\psi\vert \lambda)$ to each quantum state $\psi$ 
with $\mu(\psi\vert \lambda) > 0$ 
and an `indicator (response) function' $\xi(\psi\vert \lambda)$ to each state $\psi$ such that the Born rule is 
reproduced \cite{mon01}:
\ben
\int d\lambda\, \xi(\phi\vert \lambda)\mu(\psi\vert \lambda) &=& \vert\langle \phi\vert \psi\rangle\vert^2,\label{1}\\
\int d\lambda\, \mu(\psi\vert \lambda) &=& 1.\label{2}
\een

According to HS, an ontological model can be classified as (i) $\psi$-complete which is $\psi$-ontic, (ii) $\psi$-supplemented which is also ontic but incomplete, 
and (iii) $\psi$-epistemic which is $\psi$-incomplete. The $\psi$-complete model makes the identification $\Lambda = {\cal{CP(H)}}_{qm}$, the complex projective Hilbert 
space of quantum mechanics, and prescribes $\mu(\psi\vert \lambda) = \delta (\lambda - \psi)$. There are, however, two different ways, according to HS, in which an 
ontological model can be incomplete. It can be ontic and yet incomplete if additional variables (collectively labeled by $\omega$) are required to complete the theory, 
as in hidden variable models. In such cases, $\Lambda = (\psi, \omega)$ and the quantum state is termed $\psi$-supplemented. Another way in which the quantum state can be 
incomplete is when an ontic state $\lambda$ corresponds to two or more quantum states $\psi$ and $\phi$ corresponding to two distributions
 $\mu(\psi|\lambda)$ and $\mu(\phi|\lambda)$ over $\Lambda$ with a non-zero overlap $\Delta$. In this case, an ontic state in $\Delta$ does not encode the quantum state which can therefore 
be regarded as epistemic, i.e. mere knowledge. According to HS, Einstein favored this interpretation of the quantum state. 
 
Below, we will show how to rule out $\psi$-epistemic ontological models within the HS definition.
Now we state one of the main result as a theorem.

{\flushleft {\bf Theorem-I}} 
%$\psi$-epistemic states in ontological models that satisfy the Born probability rule given by conditions (1) and
%(2), with $\mu(\psi|\lambda) > 0$ are inconsistent with the Schr{\"o}dinger evolution.
$\psi$-epistemic ontological models that satisfy the Born probability rule given by conditions (1) and
(2), with distributions $\mu(\psi|\lambda) > 0$ in open sets $\Lambda_\psi$ for all $\psi$ are inconsistent with the Schr{\"o}dinger evolution.

{\flushleft {\em Proof.--}}
Let $\ket{\psi}$ and $\ket{\phi}$ be two distinct non-orthogonal epistemic states corresponding to an ontic state $\lambda$ in the overlap region
 $\Delta =\mu(\psi|\lambda)\cap \mu(\phi|\lambda)$ with $\Delta$ an open interval in $\Lambda$.
Consider the quantum state $\ket{\psi(t)} \in {\cal H}$ at time $t$ which satisfies the Schr{\"o}dinger equation 
\beq
i\hbar \frac{d|\psi(t)\rangle }{dt} = H|\psi(t)\rangle,
\eeq
where $H$ is the Hamiltonian of the system. Invoking the Born rule in the ontological model and putting $|\phi \rangle = |\psi(t)\rangle$ in Eqn. (\ref{1}), 
one obtains
\beq
\int d\lambda\, \xi(\psi(t)\vert \lambda)\mu(\psi(t)\vert \lambda) = 1\label{3}.
\eeq
By definition the response function satisfies 
\ben
\xi(\psi(t)|\lambda ) &=& 1 \, \forall \lambda \in \Lambda_\psi \label{xi}    
\een 
and $0$ elsewhere, where $\Lambda_\psi = \{\lambda|\mu(\psi(t)|\lambda) > 0 \}$. Note that we require $\mu(\psi|\lambda) > 0$ to 
avoid $\mu(\psi|\lambda)=0$ for some values of $\lambda \in \Delta$, because the response function $\xi(\psi|\lambda )$ need not be unity for such values 
\cite{lei}. 

Now consider two distinct quantum states  $|\phi \rangle = | \psi(t + dt) \rangle$ and $|\psi\rangle = |\psi(t) \rangle$. Then, by working to leading order in $dt$, 
we have $|\langle \psi(t+dt)|\psi(t)\rangle|^2 = 1 - 
\frac{dt^2}{\hbar^2}(\Delta H)_\psi^2$ where $(\Delta H)_\psi^2 = \langle \psi\vert (H - \langle H\rangle_\psi)^2\vert \psi\rangle$, $H$ being 
the Hamiltonian generating the unitary time evolution, and $\langle H\rangle_\psi = \langle \psi\vert H\vert \psi\rangle$. Note that for a quantum 
state to evolve in the projective Hilbert space a necessary and sufficient condition is that it should have a non-zero speed $v$, where $v$ is defined 
as $v=\frac{dD}{dt} = \frac{2 (\Delta H)_\psi}{\hbar}$ and $dD$ is the infinitesimal Fubini-Study metric over the projective Hilbert space \cite{aa,akp}. 
The infinitesimal distance as measured by the 
Fubini-Study metric on the projective Hilbert space is defined as
\beq
dD^2 = 4(1- |\langle \psi(t+dt)|\psi(t)\rangle|^2 =  \frac{4 dt^2}{\hbar^2}(\Delta H)_\psi^2. 
\eeq
Thus, the energy fluctuation $(\Delta H)_\psi $ drives the quantum state in ${\cal CP}{(\cal H})$. 
However, from Eqn.(1) we have 
\ben
&& \int d\lambda\, \xi(\psi(t+dt)\vert \lambda)\mu(\psi(t)\vert \lambda) \nonumber\\
&=& \int d\lambda\, \left[\xi(\psi\vert \lambda) + d \xi(\psi\vert \lambda) + \frac{1}{2} d^2 \xi(\psi\vert \lambda)+\cdots\right] \mu(\psi(t)\vert \lambda)\nonumber\\
&=& 1
\een
because the indicator or response function $\xi(\psi|\lambda)$ is constant over $\Lambda_\psi$ (Eqn.(\ref{xi})) and does not have any explicit $\psi$ or $t$ dependence, 
resulting in $d^n\xi(\psi\vert \lambda)=0\,\forall {n}$. This contradicts Eqn. (\ref{1}) and completes the proof.

Thus, even though Eqns (1) and (2) can reproduce the probabilistic predictions at any given time, it cannot reproduce the predictions of the Schr{\"o}dinger
time evolution at later times. This is an alternative proof of the no-go theorem for $\psi$-epistemic models. The theorem is a consequence of
the $\psi$-epistemic states having continuous Hamiltonian evolution but not the indicator or response functions. Furthermore, since the ontic states in $\Delta$ do not encode 
the epistemic states, evolution of the latter do not reflect any evolution of the former. 
%Proof of our theorem requires that the response functions are everywhere differentiable, but, there may be examples, where they need not even be 
%everywhere continuous. Then, we cannot apply Theorem-I to such models.
We should add that the proof of Theorem-1 is based on the analyticity of $\xi(\psi|\lambda)$ in the open set  $\Lambda_\psi$.  
Epistemic models that do not satisfy this condition are not covered by the theorem.

{\flushleft {\em An alternative ontological model.--}}
Ontological models are supposed to reflect closely the underlying reality that our physical theories are supposed to describe. However, quantum mechanics has been 
riddled with the measurement problem and nonlocality, features that one would like to avoid in an ontological model. We show in this letter 
that this objective can be met by (i) assigning a complex projective Hilbert space structure ${\cal{CP(H)}}$ to the ontic space in which the projective 
Hilbert space ${\cal{CP(H)}}_{qm}$ is embedded, and (ii) changing the definition of $\psi$-epistemic from the one given by HS. We will illustrate these ideas in 
greater detail below. The ontic states in this larger space will be denoted by $\vert\lambda\rangle$ and we will refer to them as ``hidden-states''. 
For simplicity, we continue to use the same notation $\Lambda$ for our ontic space as in the previous section, though our ontic space is
different from that in the HS framework, and the two ontic spaces are logically different. The most important difference is that the HS model is
based on an ontic space $\Lambda$ constructed from probabilities which satisfy the Born rule,
whereas our ontological model is based on probability amplitudes, and quantum states are averages of ``hidden'' ontic states.
 
{\flushleft {\em Alternative Definition of $\psi$-epistemic}}
The definition of epistemic states given by HS and adopted by all subsequent authors with some variations leads to contradictions with standard quantum 
mechanics, as we proved in Theorem-I above. Hence, to see if $\psi$-epistemic models can still be saved, it is necessary to change the technical definition of
$\psi$ epistemic given by HS. According to them,  the basic definition is that `$\psi$ has an ontic character if and only if a variation of $\psi$ implies a
 variation of reality and an epistemic character if and only if a variation of $\psi$ does not necessarily imply a variation of reality.' The ontic models 
satisfy this definition by having a one-to-one correspondence between $\psi$ and $\lambda$. Epistemic states must avoid such a relationship. 
One way out is to have multiple distinct quantum states compatible with the same ontic state $\lambda$, a choice made by HS. An alternative would be to 
define a quantum state $\psi$ as an average over multiple distinct ontic states $\lambda$ with a probability amplitude that can change on obtaining new 
information about the ontic state. This is a Bayesian approach which we adopt. Both these choices imply that a single `ontic state $\lambda$ does not
 encode $\psi$', and furthermore, that a single `quantum state does not parametrize the ontic states of the model at all'. A hidden variable model and
 its generalizations, on the other hand, are characterized by the ontic space $\Lambda$ parametrized by $\psi$ and supplementary variables $\omega$. We
 will avoid such an option. 

To see how our scheme works, let us first denote a basis $M$ of ${\cal{CP(H)}}_{qm}$ whose elements are constructed from the set of quantum states $\{\vert \psi\rangle\}$. Next, 
let us assume that distinct (orthogonal) elements of $\{\vert \psi\rangle\}$ correspond to non-overlapping, non-empty dense 
sets $\{\vert\lambda\rangle\} \in \Lambda ={\cal{CP(H)}}$. This requires a partitioning ${\cal{P}}$ of $\Lambda$ into subsets $\lambda$ corresponding 
to all possible distinct quantum states which cover $\Lambda$: $\emptyset \not\subset {\cal{P}},\, \bigcup \lambda = \Lambda,\, \lambda_a \cap \lambda_b = \emptyset, \, 
\lambda_a, \lambda_b \in {\cal{P}},\, a \neq b$. This means that, for every preparation procedure $P_\psi$, there is a unique quantum 
state $\vert \psi\rangle$ but a dense open set of ontic states $\lambda_\psi$ with the probability amplitude $A(\lambda \vert P_\psi)\,\,\,\,\forall \lambda \in \lambda_\psi$, 
the quantum state being an average of the ``hidden-states''  over $A(\lambda \vert P_\psi)$ (the propensity function) defined by
\beq
\vert\psi\rangle = \int_{\lambda_\psi} d\lambda  |\lambda\rangle A(\lambda \vert P_\psi) \label{psi}
\eeq
with the requirement
\beq
\langle \psi \vert \psi\rangle = \int_{\lambda_\psi} d\lambda P(\lambda \vert P_\psi) = 1,
\eeq
where we have put $\vert A(\lambda \vert P_\psi) \vert^2 = P(\lambda \vert P_\psi)$.
Thus, $P(\lambda \vert P_\psi)$ is a probability density over the ontic space. This ensures that all quantum mechanical predictions are reproduced. In particular,  
we have 
\beq
|\langle \phi|\psi \rangle|^2 = \left( \int_{\lambda_\psi \cap \lambda_\phi} d\lambda  A^*(\lambda \vert P_\phi) A(\lambda \vert P_\psi)\right)^2,
\eeq
where $|\phi \rangle$ is non-orthogonal to $\ket{\psi}$, and $\lambda_\psi \cap \lambda_\phi \neq \emptyset$.  This is the Born rule for the transition 
probability between two quantum states.
 
Note that this description not only reproduces the Born rule, it also reproduces the amplitudes for quantum transitions, which is 
not possible in the conventional ontological models. 
In the limit of the widths of the ontic state sets $\{\lambda\}$, corresponding to uncertainties of knowledge, shrinking to points, one recovers the $\psi$-complete model.
 It is in this sense that $\psi$ is incomplete and epistemic (mere knowledge) in this model. Since it is not ruled out by Theorem-I, we will refer to this model as
 a consistent ``minimalist $\psi$-epistemic'' model.

It is worth emphasizing that barring the $\psi$-complete model, in all ontological models considered so far, one obtains averages of physical observables over 
some `hidden variables', and there is no direct relationship between these and the quantum states. We have prescribed a definite relationship between a quantum state 
and an open dense set of ontic states $\ket{\lambda} \, \, \forall \lambda \in \lambda_\psi$ given by (\ref{psi}). The ontic states $\ket{\lambda}$ may thus 
be called `hidden states'. In this sense the epistemic 
quantum states are somewhat smeared descriptions of the ontic states.

{\flushleft {\em Locality.--}} We will now explain how the `minimalist $\psi$-epistemic' model is consistent with the locality principle.
The origin of the debate on nonlocality in quantum mechanics can be traced back to Einstein's observations at the 1927 Solvay Conference. Consider the case of 
a single particle wavefunction suggested by him to demonstrate that an ontic wavefunction $\psi$ describing the particle and locality are incompatible \cite{bac}.
 After passing through a small hole in a screen, the wavefunction of the particle spreads out on the other side of it in the form of a spherical wave, and is finally 
detected by a large hemispherical detector.
The wave function propagating towards the detector does not show any privileged direction. Einstein observes:
\begin{quote}
If $|\psi|^2$ were simply regarded as the probability that at a certain point a given particle is found at a given time, it could happen that {\em the same} elementary
 process produces an action in {\em two or several} places on the screen. But the interpretation, according to which the $|\psi|^2$ expresses the probability that
 {\em this} particle is found at a given point, assumes an entirely peculiar mechanism of action at a distance, which prevents the wave continuously distributed in 
space from producing an action in {\em two} places on the screen.
\end{quote}
Einstein later remarks that this `entirely peculiar mechanism of action at a distance' is in contradiction with the postulate of relativity. 

An advantage of a consistent $\psi$-epistemic ontological model is that a sudden change or collapse of the wavefunction can be interpreted as a Bayesian updating
 on receiving new information, thus avoiding nonlocality. To see this clearly, we will follow the line of argument constructed by Norsen \cite{norsen}.
 Let $A$ and $B$ be any arbitrary pair of disjoint points on the detector. The entangled state of the particle and the detector is then
 $\vert \Psi\rangle = \frac{1}{\sqrt{2}}[\vert \psi\rangle_A \vert \chi\rangle_A + \vert \psi\rangle_B \vert \chi\rangle_B ]$ where $|\psi\rangle$ and $|\chi\rangle$
 denote the particle and the detector states respectively. In the $\psi$-complete model, there is a unique $\lambda =\Psi$. In the consistent $\psi$-epistemic model, 
on the other hand, the two terms can be taken to correspond to two disjoint elements $\lambda_A,\lambda_B\in \lambda_\Psi$. 

We are now in a position to state and prove Theorem-II which addresses the question of locality.

{\flushleft {\bf Theorem-II}} In an ontological model, $\psi$-complete and locality are incompatible, while in the consistent epistemic model, 
$\psi$-epistemic and locality are compatible. 

{\flushleft {\em Proof.--}}
The probability of simultaneous detection of the particle at $A$ and $B$ in the $\psi$-ontic model is given by 
\beq
p(1_A \wedge 1_B\vert \lambda) = p(1_A\vert \lambda) p(1_B\vert 1_A, \lambda).
\eeq
The locality assumption requires that we must have  $p(1_B\vert 1_A, \lambda) = p(1_B\vert\lambda)$. Hence, using $\lambda = \Psi$, we have 
\beq
p(1_A \wedge 1_B\vert \Psi) = p(1_A\vert \Psi) p(1_B\vert \Psi) = \frac{1}{4},
\eeq
which is inconsistent with the quantum mechanical prediction that this probability vanishes. Hence, the locality assumption is false in this model.

Now consider the consistent $\psi$-epistemic model in which $\lambda = \{\lambda_A, \lambda_B\} \in \lambda_\Psi$ and $\lambda_A \cap \lambda_B = 
\emptyset$. The probability of simultaneous detection of the particle at $A$ and $B$ is
\beq
p(1_A \wedge 1_B\vert \lambda_A \wedge \lambda_B) = p(1_A\vert \lambda_A \wedge \lambda_B) p(1_B\vert 1_A, \lambda_A \wedge \lambda_B).
\eeq
In this model the locality condition requires $p(1_A\vert \lambda_A \wedge \lambda_B) = p(1_A\vert \lambda_A)$ and $p(1_B\vert 1_A, \lambda_A \wedge \lambda_B) =
 p(1_B\vert \lambda_A) = 0$. Hence, 
\beq
p(1_A \wedge 1_B\vert \lambda_A \wedge \lambda_B) = p(1_A\vert \lambda_A) p(1_B\vert \lambda_A) = 0
\eeq
which is consistent with the quantum mechanical prediction. Hence, this model is compatible with the locality assumption. This completes the proof of the theorem.

We may remark that the ``hidden states'' do play a role in the situation considered by Einstein. Indeed we can say that one of these states is revealed by the measurement. 
The spherical wavefunction $\psi$ is an average over these states with a uniform probability amplitude, each point on the sphere corresponding to an ontic
state in the dense subset $\lambda_\psi$ which represents the uncertainties on preparation of the state. When a spot appears on the detector, it 
reveals the corresponding ontic state--thereby one can say that the measurement removes the uncertainties. The ``hidden states'' introduced in this paper 
also clearly demarcates the difference with the hidden variables which lie hidden forever.

{\flushleft {\em Concluding remarks.--}} We have proved that $\psi$-epistemic ontological models based on positive definite probability distributions 
and everywhere differentiable response functions satisfying the Born rule specified by conditions (1) and (2) are inconsistent with Schr\"{o}dinger evolution (Theorem-I). 
Thus, the ontological models 
with $\psi$-epistemic wavefunctions, though they can reproduce measurement results at a fixed time, are silent about dynamical aspects. This impelled us to look 
for an alternative ontological model which can accommodate an epistemic interpretation of the quantum state. An epistemic interpretation is preferable because 
it can do away with many conundrums of quantum theory such as measurement and nonlocality. Accordingly, we postulate that the ontic space is a 
complex projective Hilbert space $CPH$ that embeds $CPH_{qm}$ and that is partitioned into disjoint open dense sets. The quantum states appear as averages over these ``hidden states'' in the larger ontic space. This makes quantum mechanics a somewhat smeared but fairly close description of the underlying reality. 
Within this framework, the epistemic interpretation is shown to be consistent with locality (Theorem-II). We believe that although the HS definition of 
epistemic states is inspired by Einstein's views, it is our definition that achieves his objective. We hope that the minimalist $\psi$-epistemic model 
presented here can provide new insights to the nature of quantum states.

{\flushleft {\em Acknowledgements.--}}
AKR and PG thank the Quantum Information and Computation (QIC) group in HRI, Allahabad for hosting their stay and providing excellent support. PG also thanks the National  
Academy of Sciences, India for the grant of a Senior Scientist Platinum Jubilee Fellowship that enabled this work to be undertaken. AKP thanks M. K. Patra for useful remarks.
Finally, all the authors thank A. Sudbery and M. Leifer for helpful critical comments.


\begin{thebibliography}{999}

\bibitem{bell} J.\ S.\ Bell, {\em Physics} (Long Island City, N.Y.) {\bf 1}, 195 (1964).

\bibitem{bell01} J.\ S.\ Bell, {\em Rev. Mod. Phys.} {\bf 38}, 447 (1966).

\bibitem{bell02} J.\ S.\ Bell, {\em Speakable and Unspeakable in Quantum Mechanics}, 
Cambridge University Press, England, Cambridge (2004).

\bibitem{pbr} M.\ F.\ Pusey, J.\ Barrett, and T.\ Rudolph, {\em Nature Phys.} {\bf 8}, 476 (2012).

\bibitem{cr} R.\ Collbeck and R.\ Renner, {\em Phys. Rev. Lett.} {\bf 108}, 150402 (2012).

\bibitem{pg}  P.\ G.\ Lewis, D.\ Jennings, J.\ Barrett, and T.\ Rudolph, {\em Phys. Rev. Lett.} {\bf 109}, 150404 (2012).

\bibitem{patra} M.\ K.\ Patra, S.\ Pironio, and S.\ Massar, 
{\em Phys. Rev. Lett.} {\bf 111}, 090402 (2013).

\bibitem{rws} R.\ W.\ Spekkens, {\em Phys. Rev. A} {\bf 71}, 052108 (2005).

\bibitem{rws1} R.\ W.\ Spekkens, {\em Phys. Rev. A} {\bf 75}, 032110 (2007).

\bibitem{sf} M.\ Schlosshauer and A.\ Fine, {\em Phys. Rev. Lett.} {\bf 108}, 260404 (2012).

\bibitem{lei} M.\ S.\ Leifer and O.\ J.\ E.\ Maroney, {\em Phys. Rev. Lett.} {\bf 110}, 120401 (2013).

\bibitem{hardy} L.\ Hardy, {\em Int. J. Mod. Phys. B} {\bf 27}, 1345012 (2013).

\bibitem{gr} G.\ Ghirardi and R.\ Romano, {\em Phys. Rev. Lett.} {\bf 110}, 170404 (2013).

\bibitem{fuchs} C.\ A.\ Fuchs, N.\ D.\ Mermin, and R.\ Schak, arXiv: quant-ph/1311.5253.
 
\bibitem{hs}  N.\ Harrigan and R.\ W.\ Spekkens, {\em Found. Phys.} {\bf 40}, 125 (2010).

\bibitem{mon01} A.\  Montina, {\em Phys. Rev. A} {\bf 77}, 022104 (2008).

\bibitem{aa} J. Anandan and Y. Aharonov, Phys. Rev. Lett. {\bf 65}, 1697 (1990).

\bibitem{akp} A. K. Pati, Phys. Lett. A. {\bf 159}, 105 (1991). 

\bibitem{bac} G.\  Bacciagaluppi, A.\ Valentini: {\em Quantum Theory at the Crossroads: Reconsidering the 1927 Solvay Conference}, pp. 485-487, Cambridge Univ. Press, Cambridge (2009); arXiv: quant-ph/0609184.

\bibitem{norsen} T.\ Norsen, {\em Am. J. Phys.} {\bf 73}, 164 (2005).


%  12. O. J. E. Maroney, arXiv:1207.6906.
 
%  16. N. Harrigan and T. Rudolph, arXiv:0709.4266.
%  17. N. D. Mermin, Rev. Mod. Phys. 65, 803 (1993).
%  20. D. A. Meyer, Phys. Rev. Lett. 83, 3751 (1999); A. Kent, ibid. 83, 3755 (1999); R. K. Clifton and A. Kent, Proc. R. Soc. A 456, 2101 (2000); J. Barrett and A. Kent, Stud. Hist. Phil. Mod. Phys. 35, 151 (2004); R. Hermens, ibid. 42, 214 (2011).
%  21. E. Schrödinger, Proc. Cambridge Philos. Soc. 32, 446 (1936); L. P. Hughston, R. Jozsa, and W. K. Wootters, Phys. Lett. A 183, 14 (1993).

  
\end{thebibliography}
\end{document}